\documentclass[epsf,usegraphicx]{mn2e}

\usepackage{xspace}
\usepackage{amsfonts}
\usepackage{pifont}

\title[The long term behaviour of AM CVn systems]
{The long term optical behaviour of helium accreting AM CVn binaries}

\author[]
{Gavin Ramsay$^{1}$, Thomas Barclay$^{1,2}$,  
Danny Steeghs$^{3}$, Peter J. Wheatley$^{3}$, \and Pasi Hakala$^{4}$
Iwona Kotko$^{5}$ Simon Rosen$^{6}$\\
$^{1}$Armagh Observatory, College Hill, Armagh, BT61 9DG\\
$^{2}$Mullard Space Science Laboratory, University College London,
Holmbury St. Mary, Dorking, Surrey, RH5 6NT\\
$^{3}$Department of Physics, University of Warwick, Coventry, CV4 7AL\\
$^{4}$Finnish Centre for Astronomy with ESO (FINCA) , University of Turku,
V\"{a}is\"{a}l \"{a}ntie 20, FI-21500 PIIKKI\"{O}, Finland\\
$^{5}$Astronomical Observatory, Jagiellonian University, Cracow, Poland\\
$^{6}$Department of Physics and Astronomy, University of Leicester, 
University Road, Leicester LE1 7RH\\
}

\date{Accepted MNRAS 2010 Sept 29.}

\begin{document}
\outer\def\gtae {$\buildrel {\lower3pt\hbox{$>$}} \over 
{\lower2pt\hbox{$\sim$}} $}
\outer\def\ltae {$\buildrel {\lower3pt\hbox{$<$}} \over 
{\lower2pt\hbox{$\sim$}} $}
\newcommand{\ergscm} {ergs s$^{-1}$ cm$^{-2}$}
\newcommand{\ergss} {ergs s$^{-1}$}
\newcommand{\ergsd} {ergs s$^{-1}$ $d^{2}_{100}$}
\newcommand{\pcmsq} {cm$^{-2}$}
\newcommand{\ros} {\sl ROSAT}
\newcommand{\chan} {\sl Chandra}
\newcommand{\xmm} {\sl XMM-Newton}
\newcommand{\swift} {\sl Swift}
\def\rchi{{${\chi}_{\nu}^{2}$}}
\newcommand{\Msun} {$M_{\odot}$}
\newcommand{\Mwd} {$M_{wd}$}
\def\Mdot{\hbox{$\dot M$}}
\def\mdot{\hbox{$\dot m$}}
\newcommand{\teff}{\ensuremath{T_{\mathrm{eff}}}\xspace}
\newcommand{\tickYes}{\checkmark}
\newcommand{\tickNo}{\hspace{1pt}\ding{55}}

\maketitle

\begin{abstract}

  We present the results of a two and a half year optical photometric
  monitoring programme covering 16 AM CVn binaries using the Liverpool
  Telescope on La Palma. We detected outbursts in seven systems, one
  of which (SDSS J0129) was seen in outburst for the first time. Our
  study coupled with existing data shows that $\sim$1/3 of these
  helium-rich accreting compact binaries show outbursts. The orbital
  period of the outbursting systems lie in the range 24--44 mins and
  is remarkably consistent with disk-instability predictions. The
  characteristics of the outbursts seem to be broadly correlated with
  their orbital period (and hence mass transfer rate). Systems which
  have short periods ($<$30 min) tend to exhibit outbursts lasting
  1--2 weeks and often show a distinct `dip' in flux shortly after the
  on-set of the burst. We explore the nature of these dips which are
  also seen in the near-UV.  The longer period bursters show higher
  amplitude events (5 mag) that can last several months. We have made
  simulations to estimate how many outbursts we are likely to have
  missed.

\end{abstract}


\section{Introduction}

AM CVn systems are accreting binaries consisting of a white dwarf
accretor and a degenerate (or semi-degenerate) secondary. They are
notable for at least two reasons: they have an extremely short orbital
period ($P_{orb}<$ 70 mins) and are almost entirely hydrogen deficient
(see Solheim 2010 for a recent review). There are currently 27 known
AM CVn systems of which half a dozen have been found to exhibit at
least one outburst, where they increase in brightness typically by
3--4 magnitudes. Indeed, some systems were discovered through
supernovae surveys (eg V406 Hya =2003aw, Wood-Vasey et al. 2003). It is
thought that the origin of these outbursts is similar to those seen in
hydrogen accreting dwarf novae (eg Osaki 1996). Much observational and
theoretical work has been done in trying to understand the origin and
nature of the outbursts in the hydrogen accreting systems (eg Lasota
2001). In contrast, our understanding of the cause of the outbursts
seen in AM CVn systems is much more limited.

On the observational side, this is largely due to the lack of
information on their long term behaviour. On the theoretical side,
progress has been more limited compared to the work on hydrogen
accreting systems. However, the work of Smak (1983), who described a
thermal instability model, Whitehurst (1988) and Tsugawa \& Osaki
(1997) who develop tidal instability models and Lasota, Dubus \& Kruk
(2008) and Kotko, Lasota \& Dubus (2010) who take into account
irradiation effects, are some notable advances in modelling hydrogen
deficient accreting binaries.

\begin{table*}
\begin{center}
\begin{tabular}{lrccrrrr}
\hline
Source & Period &LT      &  LT Range & LT mean & LT rms & Outbursting  &Reference \\
       & (mins) &Target? &  g (mag)  & g (mag)  & g (mag)  &  System?       &          \\
\hline
HM Cnc                   & 5.4  &\tickNo  &        & &     &N &  [1]\\
V407 Vul                 & 9.5  &\tickNo  &         & &    &N &  [2]\\
ES Cet                   & 10.4 &\tickYes & 16.5--16.8 & 16.66 & 0.09 &N &  [3]\\
SDSS J190817.07+394036.4 & 15.6 &\tickNo  &        & &     &N &  [4]\\
AM CVn                   & 17.1 &\tickNo  &         & &    &N &  [5]\\
HP Lib                   & 18.4 &\tickYes & 13.6--13.7 & 13.51 & 0.02& N &  [6]\\
CR Boo                   & 24.5 &\tickYes & 13.8--17.0 & 15.00 & 1.00 &Y &  [7]\\
KL Dra                   & 25.0 &\tickYes & 16.0--19.6 & 17.68 & 1.19 &Y &  [8]\\
PTF1 J0719+4858          & 26.8 &\tickNo  &        & &     &Y &  [25]\\ 
V803 Cen                 & 26.9 &\tickNo  &         & &    &Y &  [9]\\
SDSS J092638.71+362402.4 & 28.3 &\tickYes & 16.6--19.6 & 19.31 & 0.52 &Y &  [10]\\
CP Eri                   & 28.4 &\tickYes & 16.2--20.2 & 19.30 & 1.33& Y &  [11]\\
V406 Hya (2003aw)        & 33.8 &\tickYes & 14.5--19.7 & 18.89 & 1.34 &Y &  [12]\\
2QZ J142701.6-012310     & 36.6 &\tickYes & 20.0--20.5 &20.35 & 0.18&Y &  [13]\\
SDSS J012940.05+384210.4 &:37   &\tickYes & 14.5--20.0 &19.31 & 1.31 &Y &  [14, 15, 16] \\
SDSS J124058.03-015919.2 & 37.4 &\tickYes & 19.0--19.8 &19.62& 0.17&Y &  [16, 17] \\
SDSS J080449.49+161624.8 & 44.5 &\tickYes & 17.8--19.0 &18.54& 0.30&Y &  [18]\\
SDSS J141118.31+481257.6 & :46  &\tickYes & 19.4--19.7 &19.58& 0.09&N &  [14]\\
GP Com                   & 46.6 &\tickYes & 15.9--16.3 &16.22 & 0.05&N &  [19]\\
SDSS J090221.35+381941.9 & 48.3 &\tickNo  &       & &     &N &  [20]\\
SDSS J155252.48+320150.9 & 56.3 &\tickYes & 20.2--20.6 &20.44& 0.10&N &  [21]\\
V396 Hya                 & 65.1 &\tickNo  &        & &    &N &  [22]\\
SDSS J120841.96+355025.2 &      &\tickYes & 18.9--19.4 &19.09&0.07&N &  [23]\\
SDSS J152509.57+360054.5 &      &\tickYes & 19.8--20.2 &19.91 & 0.11&N &  [20]\\
SDSS J164228.06+193410.0 &      &\tickNo  &        & &     &N &  [20]\\
SDSS J172102.48+273301.2 &      &\tickYes & 20.4--20.7 & 20.53&0.17&N &  [20]\\  
SDSS J204739.40+000840.1 &      &\tickYes & 17.0--17.4 &17.13&0.03&Y &  [24]\\ 
\hline
\end{tabular}
\end{center}
\caption{The currently known AM CVn binaries. We show their period
  which is either the orbital period (most reliably determined from
  spectroscopic observations) or the dominant photometric period which
  is typically within a few percent of the orbital period. A colon implies the
  period is approximate, while the lower five systems do not have
  known periods. We show by means of a tick or a cross whether the
  source is included our Liverpool Telescope survey and the observed
  range, the mean mag and the rms of the light curve 
 in the $g$ band magnitude. We also show through a `Y' or
  a `No' whether the source has been seen in outburst, either in
  these LT observations or through previous work.  References: [1]
  Ramsay, Cropper \& Hakala (2002), [2] Ramsay et al. (2000), [3]
  Espaillat et al. (2005), [4] Fontaine et al. (2011), [5] Nelemans,
  Steeghs \& Groot (2001), [6] O'Donoghue et al. (1994) [7] Wood et al.
  (1987), [8] Wood et al. (2002), [9] Kato et al. (2004), [10]
  Copperwheat et al. (2011), [11] Abbott et al. (1992), [12] Wood-Vasey
  et al. (2003), [13] Woudt, Warner \& Rykoff (2005), [14] Anderson et
  al (2005), [15] Barclay et al. (2009), [16] Shears et al. (2011), [17]
  Roelofs et al. (2005), [18] Roelofs et al. (2009), [19] Nather,
  Robinson \& Stover (1981), [20] Rau et al. (2010), [21] Roelofs et al.
  (2007a), [22] Ruiz et al. (2001), [23] Anderson et al. (2008), [24]
  Prieto et al. (2006), [25] Levitan et al. (2011).}
\label{sources}
\end{table*}

Accretion discs in AM CVn binaries can be thermally stable in a hot
state -- the temperature of the disc is always greater than the
ionisation temperature of helium -- and can also be thermally stable
in a low state -- the temperature of the disc is always lower than the
ionisation state of helium. Systems which lie between these states are
expected to be unstable. It is thought that the longer period systems
(40 min $< P_{orb} <$ 70 min) lie in the cool stable region
and therefore do not show large optical photometric variations. Little
is known about the long-term photometric behaviour of most AM CVn
systems, in particular the fainter half of the sample which has been
discovered in recent years. These systems are generally beyond the
reach of amateur variable star networks (although see Shears et al.
2011), and coverage on larger telescopes has been patchy.  To better
characterise the properties of AM CVn systems and obtain a better
understanding of the physics of the outbursts, we began a monitoring
programme using the Liverpool Telescope in Feb 2009 and ended in June
2011.

\section{Liverpool Telescope observations}

The Liverpool Telescope is a 2m robotic telescope sited on the island
of La Palma in the Canaries and has a suite of instruments including
the RATCAM imager (Steele 2004). Our goal was to obtain one $g$ band
image per source once per week for the first year and once every 5
days thereafter.  Our target list (see Table \ref{sources}) include 18
out of the 27 known AM CVn systems. 

\begin{figure*}
\begin{center}
\setlength{\unitlength}{1cm}
\begin{picture}(16,12.8)
\put(-0.5,0.5){\includegraphics{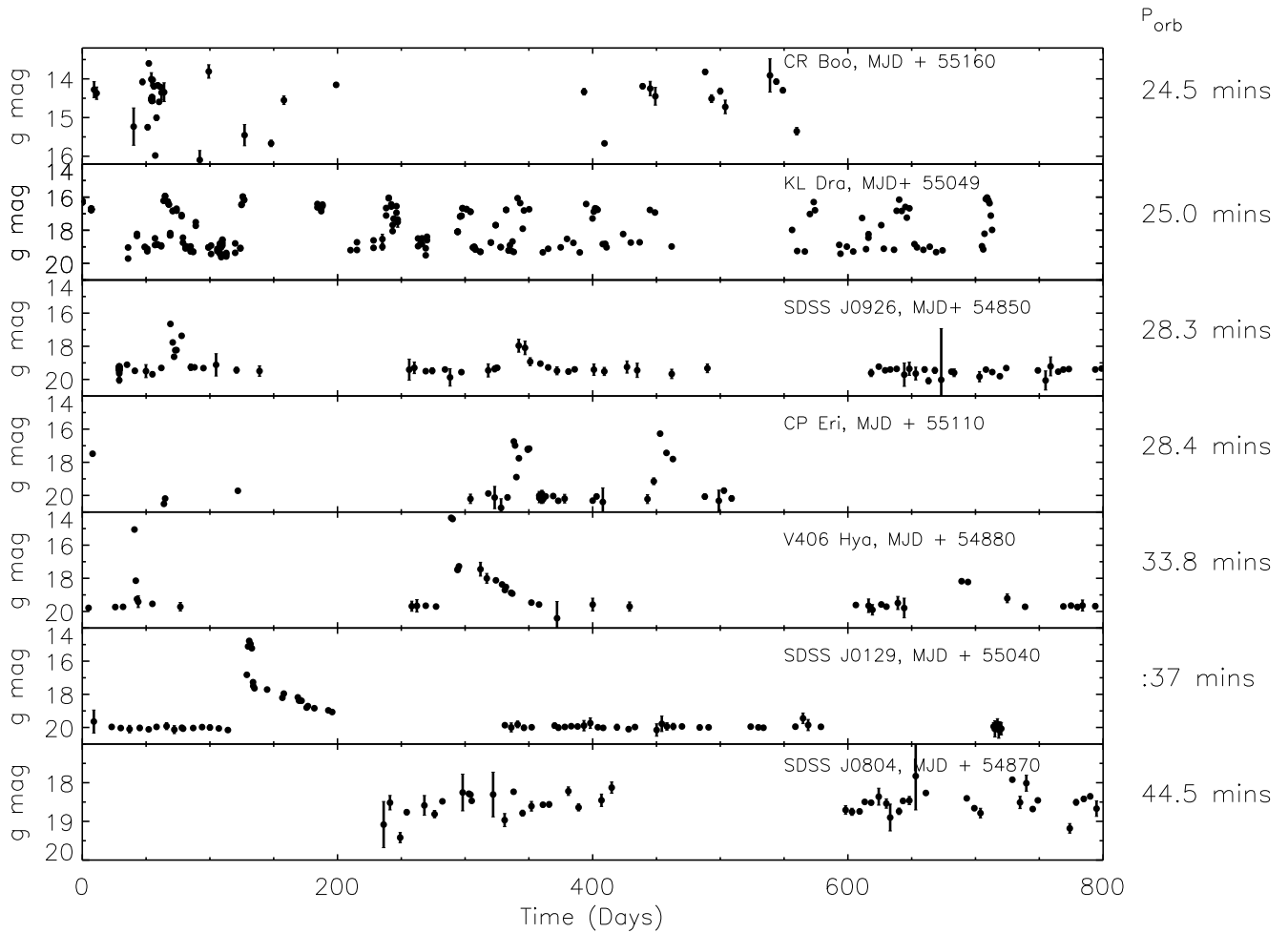}}
\end{picture}
\end{center}
\caption{The light curves of AM CVn systems obtained using the
  Liverpool Telescope which showed an outburst. We show the start date
of the observations in the top right of each panel and the orbital
period of each system towards the right.}
\label{outbursts}
\end{figure*}

Exposure times ranged from 5 sec in the brightest source (GP Com) to
140 sec in the faintest (eg SDSS J0129). Observations were made
  in a range of transparencies, lunar brightness and airmass.  Images
were typically downloaded within a few days of them being taken and
have already been bias subtracted and flat-fielded in an automatic
way. We selected two comparison stars and used the photometry analysis
tool {\tt autophotom} which is incorporated in the package {\tt Gaia}
to determine the relative brightness of our source with respect to the
comparisons\footnote{Both packages are part of the STARLINK suite of
  software which can be downloaded from
  http://starlink.jach.hawaii.edu/starlink} . Some fields have
  low stellar density which meant that the comparison stars were as
  faint as $g\sim$19. We derived a differential light curve between
  the comparison stars used for each AM CVn system. The mean rms of
  these light curves was 0.04 mag when both comparison stars were
  brighter than $g$=18, and 0.11 mag when both stars were fainter than
  $g$=18. Since our aim is to detect variations greater than 1 mag,
  this is sufficient for our purposes.  To place our results on to
the standard $g$ band system, we used SDSS photometry of stars in the
immediate field when available. For other sources we obtained images
of the field (along with standard star fields) using the INT on La
Palma.

\begin{figure}
\begin{center}
\setlength{\unitlength}{1cm}
\begin{picture}(8,13)
\put(-0.5,-0.5){\includegraphics{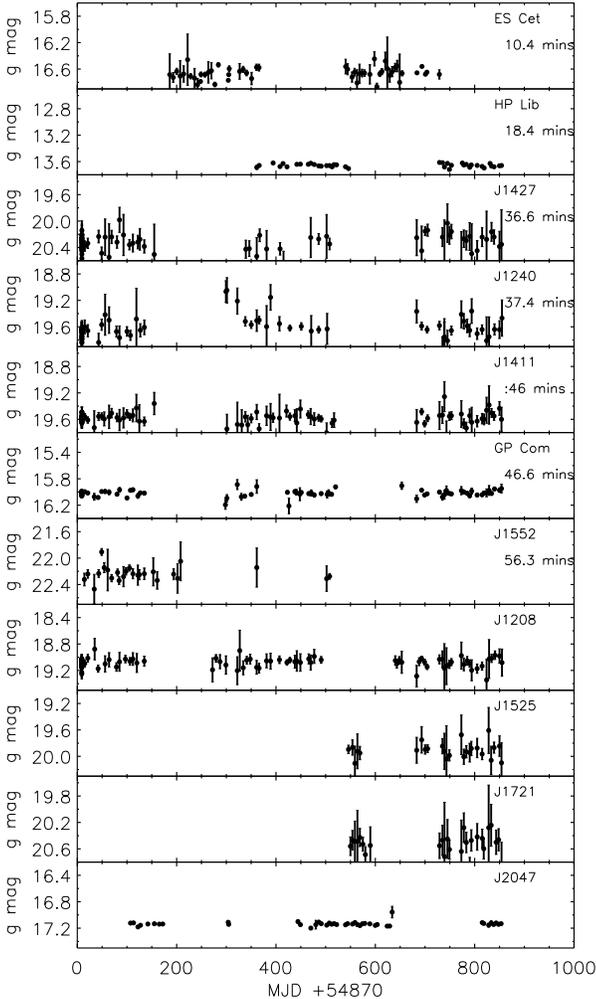}}
\end{picture}
\end{center}
\caption{The light curves obtained using the Liverpool Telescope of 
those systems which did not show an outburst. We show the orbital
period of each system towards the right hand side of each panel. The
orbital period is not known for the lower four systems.}
\label{nooutbursts}
\end{figure}

We show the light curve of each system in Figure \ref{outbursts}
  (for those showing outbursts) and in Figure \ref{nooutbursts} (for
  those which did not). The mean magnitude and range in brightness of
  our sources as derived using LT data is shown in Table
  \ref{sources}, along with the rms of the light curve (see also
  Figure \ref{rms}).

\section{Results}

\subsection{Overview}

\begin{figure}
\begin{center}
\setlength{\unitlength}{1cm}
\begin{picture}(8,7)
\put(-1.5,0){\includegraphics{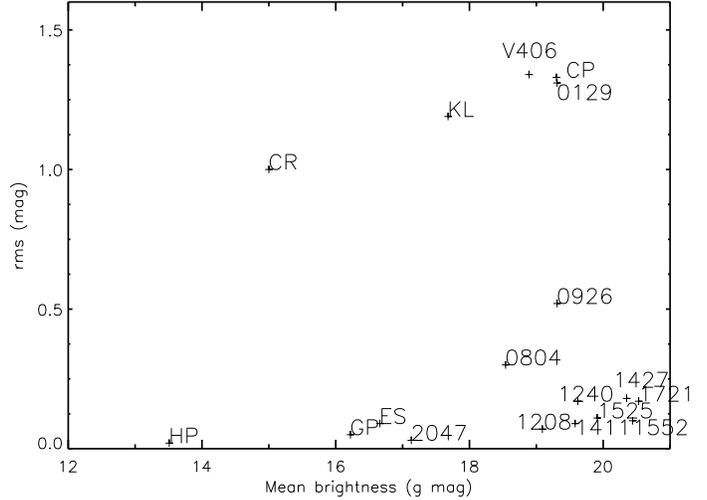}}
\end{picture}
\end{center}
\caption{Using our LT data we show the mean $g$ band brightness of our sources 
as a function of the root mean squared of the light curve.}
\label{rms}
\end{figure}

Outbursts from hydrogen accreting cataclysmic variables (CVs) show
  amplitudes of at least one mag (eg Koto, Lasota \& Dubus 2010).  We
  therefore define an AM CVn system in outburst as that which shows
  light curve variations greater than 1 mag.  Of the 18 AM CVn systems
  in our LT survey, seven were found to show at least one outburst (CR
  Boo, KL Dra, SDSS J0926, CP Eri, V406 Hya, SDSS J0129 and SDSS
  J0804). The LT data of SDSS J0129 were the first to show this system
  in outburst (Barclay et al. 2009).  The other sources all had rms
  variations less than 0.2 mag (Figure \ref{rms}).  Two systems which
  have previously been seen in outburst, (2QZ J1427 and SDSS J1240)
  were not seen in outburst in our LT data (although it is possible we
  have detected a decline from outburst in SDSS J1240 at
  MJD$\sim$55170).  Combining our results with those in the
  literature, we find that out of the 27 AM CVn systems currently
  known, 10 show outbursts, and they fall into a narrow range of
  orbital periods, covering 24.5--44.5 min.

\begin{figure}
\begin{center}
\setlength{\unitlength}{1cm}
\begin{picture}(8,13)
\put(-0.5,-0.5){\includegraphics{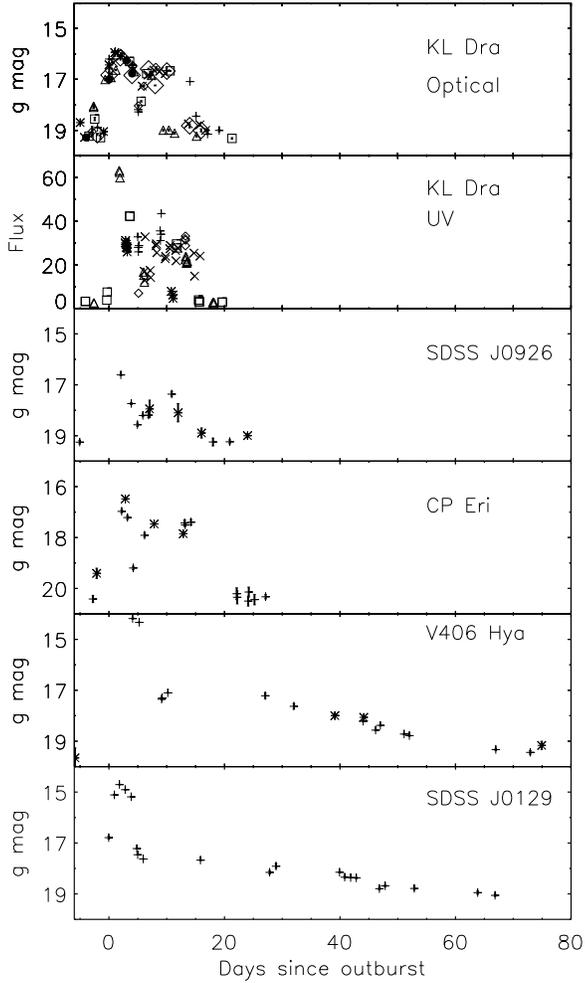}}
\end{picture}
\end{center}
\caption{We show typical outburst profile of outbursting AM CVn
  systems. Different symbols have been used for different
  outbursts. The UV data from KL Dra has been obtained using the {\sl
    Swift} satellite.}
\label{mean}
\end{figure}

  To get an overview of the outburst characteristics of each source we
  show in Figure \ref{mean} the outburst profile of each source (with
  the exception of CR Boo and SDSS J0804 which show erratic
  outbursts). For sources which have shown more than one outburst we
  over-lay data from different outbursts.  The shape of the outburst
  light curves can be split into three broad classes: relatively long
  duration events such as those seen in V406 Hya and SDSS J0129;
  shorter duration bursts such as SDSS J0926, CP Eri and KL Dra
  (Ramsay et al. 2010), and rapid transitions from bright to faint
  states as seen in CR Boo and SDSS J0804.  We show the typical
  duration, amplitude and duty cycle (the fraction of observations
  which showed the source in outburst) of each source in Table
  \ref{burst-typical}.  For systems with orbital periods between
  33--37 mins, there is a tendancy for the outbursts to have longer
  durations and higher amplitude than the shorter period systems. 

\subsection{Likelihood of missing outbursts}

\begin{figure}
\begin{center}
\setlength{\unitlength}{1cm}
\begin{picture}(8,11)
\put(-1,-2.2){\includegraphics{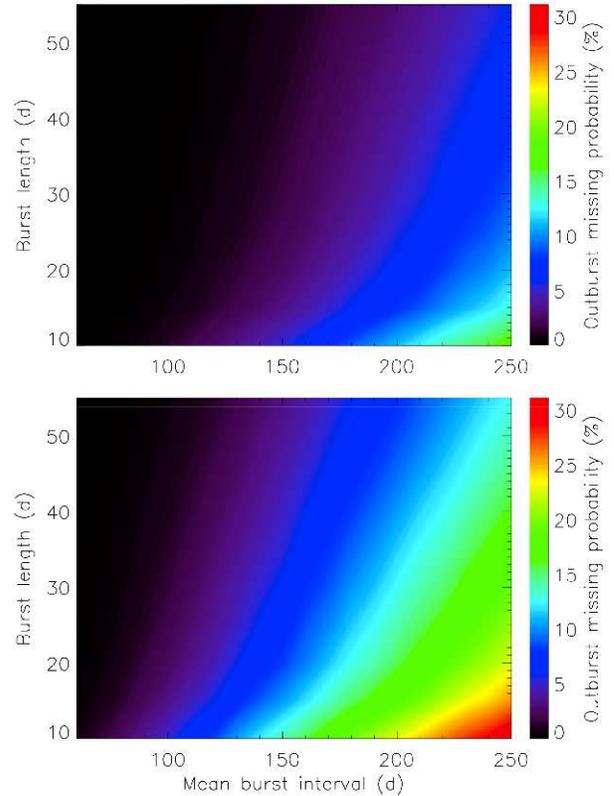}}
\end{picture}
\end{center}
\caption{We show the probability of missing an outburst as a function
  of mean outburst interval and outburst length. In the top panel we
  derive the probability for GP Com and in the lower panel J2047.}
\label{sims}
\end{figure}

We have carried out simulations in order to estimate
  characteristic probabilities for missing an outburst given the
  typical data sampling. In particular, we have created synthetic
  light curves with a range of outburst lenghts and separation
  intervals.  This was done by assuming an exponential distribution
  for inter-outburst times and a uniform distibution for outburst
  lengths. A range of different means were used for the inter-outburst
  exponential distributions. Figure \ref{sims} shows two
  representative cases. We have plotted a 2D probability distribution
  of missing outbursts from GP Com (top) and J2047 (bottom) as a
  function of mean inter-outburst duration versus the outburst
  length. GP Com and J2047 represent two cases where we did not detect
  an outburst. For both systems, the probability of us missing an
  outburst is less than 10 percent when the burst is interval is less
  than 150 days. For longer burst intervals the probability of us
  missing a burst reaches 30 percent in the case of J2047 when the
  burst length is shorter than 2 weeks.  

  We also examined the light curve of J0926 which shows two outbursts
  although we may well have missed additional bursts. We performed a
  similar set of simulations although on this occasion we define the
  outburst length as being 10 days. We find the number of missed
  outbursts does not depend on the outburst frequency - we expect to
  miss 47 percent of outbursts if the outburst frequency is between 60
  and 400 days. Our data of J0926 unevenly cover 800 days and our
  similations predict three outbursts over this timescale where we
  detected two.

  We now proceed and discuss individual outbursting systems in more
  detail.

\begin{table}
\begin{center}
\begin{tabular}{lrrrr}
\hline
Source   & Orbital           & Duration & Amplitude & Duty  \\
             & Period  (mins) & (days)      & (mag)        & Cycle \\
\hline
CR Boo & 24.5 & & 2.3 & 70$\%$ \\
KL Dra & 25.0 & 10 & 3.3 & 15$\% $\\ 
J0929  & 28.3 & 10 & 2.6 & 8$\% $\\ 
CP Eri & 28.4 & 15 & 4.0 & 27$\% $\\ 
V406 Hya & 33.8 & 50 & 5.2 & 27$\% $\\ 
J0129 & :37 & 50 & 4.2 & 22$\% $\\ 
J0804 & 44.5 & & 1.2 & 8$\% $\\ 
\hline
\end{tabular}
\end{center}
\caption{The typical duration, amplitude and duty cycle (the
  percentage of observations for which the source was seen in
  outburst) of those AM CVn systems seen in outburst in our LT data.
CR Boo and SDSS J0804 showed erratic rather than regular outbursts.}
\label{burst-typical}
\end{table}

\subsection{CR Boo}

CR Boo shows marked changes in its intensity with an amplitude of 3
mag. It is either brighter than $g$=14.5, or fainter than $g$=16.5
  for 90 percent of time and is brighter than $g$=14.5 for 70 percent
  of the epochs it was observed using the LT.  However, our sampling
is not frequent enough to determine if there is an underlying periodic
or quasi-periodic signal. However, observations covering a timescale
of 3 years show evidence for a quasi periodic transition between
bright and faint states of 46.3 days (Kato et al. 2001). On the other
hand, Patterson et al. (1997) found a prominent modulation ($\sim$1
mag) on a period of $\sim$19--22 hrs (which our data are unable to
sample).

\subsection{KL Dra}

\begin{figure}
\begin{center}
\setlength{\unitlength}{1cm}
\begin{picture}(8,8.5)
\put(-0.5,-2){\includegraphics{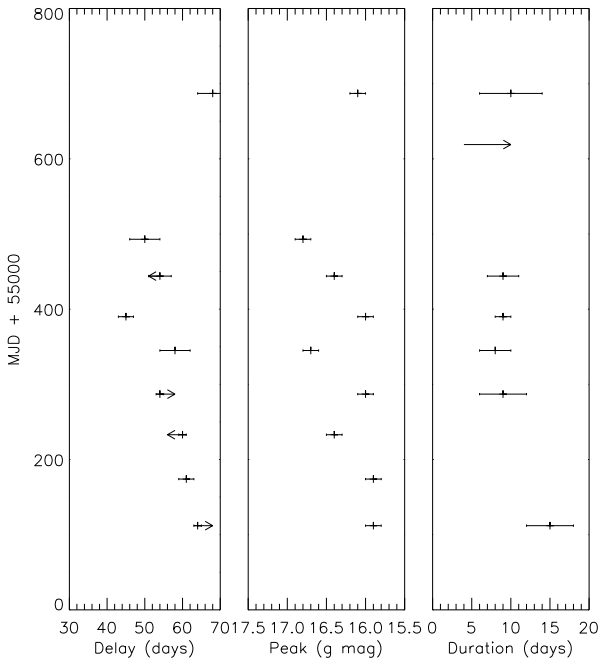}}
\end{picture}
\end{center}
\caption{KL Dra: In the left hand panel we show the delay between the
  burst peaking at the date shown on the y-axis (MJD-55000) and the
  previous one. In the middle panel we show the $g$ mag peak
  brightness of the burst, while in the right hand panel we show the
  duration of the burst.}
\label{kldra}
\end{figure}

The first nine months of our LT observations of KL Dra have been
reported in Ramsay et al. (2010). They showed an outburst every
$\sim$60 days  (or roughly six per year), making KL Dra the best
example of a helium accreting dwarf nova. Since then KL Dra has
continued to show regular outbursts. We have determined the recurrence
time between successive outbursts and their peak brightness. Although
there is some uncertainty in determining both parameters since the
cadence of our observations is variable, it is clear that the
recurrence time shortened from 60 days at the beginning of our
observations, to $\sim$45 days after a further five outbursts (Figure
\ref{kldra}). When KL Dra became visible again the recurrence time had
increased to $\sim$65 days.  There is a hint that the peak brightness
in the optical band followed the trend seen in the recurrence time.

Although our coverage is such that the duration of the outbursts is
not well defined, the outburst with the best coverage reported by
Ramsay et al. (2010) lasted 17 days. However, for the cycle which had a
recurrence time of 44 days, the duration of the outburst prior to this
lasted only 9 days. The typical duty cycle for the system being in the
outburst state is 15 percent. The fact that there is a significant
spread in the outburst recurrence time is entirely consistent with
that seen in the hydrogen accreting dwarf novae, which can show a
spread of up to a factor of 4 (cf Cannizzo, Shafter \& Wheeler
1988). We note that the {\sl Kepler} observations of the dwarf nova
V344 Lyr shows a shortening in the recurrence time of the outbursts
immediately after a super-outburst, which then appears to lengthen and
shorten in the lead up to the next super-outburst (Cannizzo et al.
2010).

\subsection{SDSS J0926}

Observations made using the Catalina Real-Time Transient Survey showed
five outbursts (of amplitude $\sim$3 mag) of SDSS J0926 over nearly 6
years of observations (Copperwheat et al. 2011). (Our simulations
  suggest that roughly eight outbursts would be expected over 6
  years). The sampling is such that the source was observed, at best,
once every 20 days or so. Our observations made using the LT show one
clear outburst with an amplitude of $\sim$2.5 mag and a duration
$\sim$16 days, with a second outburst for which we caught the decline
phase (at MJD$\sim$55192, cf Figure \ref{outbursts}). Our data
  show SDSS J0926 was brighter than $g$=18.4 mag for 8 percent of our
  observations. We note the presence of a brightening 8 days after
the initial outburst -- this maybe related to the dip seen in both CP
Eri and KL Dra (Figure \ref{mean}. We will discuss these dips further
in \S 4.

\subsection{CP Eri}

We have detected CP Eri in outburst on three occasions (although in
the first there is only one data point where our sampling was rather
sparse). The amplitude of the outbursts is $\sim$4 mag, while the rise
time from the faint state to the high state was less than 5 days in
the second outburst and less than 10 in the third. Our data show
  CP Eri was brighter than $g$=18.0 mag for 27 percent of our
  observations.  We note a short duration drop in the intensity of CP
Eri 2 days after the on-set of the second burst (cf Figure \ref{mean} and
\S \ref{dips}). The third outburst took place $\sim$114 days after the
second. The only other recorded instances of CP Eri showing outbursts
is Luyten \& Haro (1959) who found it $\sim$2.5 mag brighter than on
the POSS images and Abbot et al. (1992) observed it at one epoch when
it was brighter still.

\subsection{V406 Hya}
\label{v406hya}

V406 Hya was originally thought to be a supernova (2003aw) but
subsequent spectroscopic observations showed it was an AM CVn
system. Followup photometric observations showed that V406 Hya
exhibits outbursts which can be up to 5 mag in amplitude and have
durations lasting two or three months (Woudt \& Warner 2003, Nogami et
al 2004).  Our observations using the LT showed one long duration
outburst, which lasted more than 60 days and had an amplitude of 4
mag. We also observed an increase in brightness at MJD$\sim$56170
which could be the decline from an outburst which was not detected
(Figure 1).  Although our observations did not have high cadence, they
indicate that the rise to maximum took place on a timescale $<$12
days, and showed a very rapid initial decline from maximum (4 days)
followed by a lengthy decline phase. Our data show V406 Hya was
  brighter than $g$=18.5 mag for 27 percent of our observations. The
characteristics of this outburst are very similar to that found in
Nogami et al. (2004). Our observations also show a very short duration
(less than 10 days), high amplitude outburst (maximum at MJD=54920).

\subsection{SDSS J0129}

SDSS J0129 was seen in outburst on Dec 3rd 2009 when it was seen 3 mag
brighter than it was 14 days earlier. It went on to reach maximum
brightness after a further two days when it was 5 mag brighter than
found in quiescence (Barclay et al. 2009).  The shape of the light
curve is very similar to V406 Hya, showing a very rapid initial
decline from maximum brightness (less than a day) and an extended
tail. Our data show SDSS J0129 was brighter than $g$=18.5 mag for
  22 percent of our observations.

Shears et al. (2011) have recently reported the detection of the same
outburst as reported here. Moreover they detected super-humps during
the outburst which had a period of 38 min. Since the super-hump 
period is typically longer than the orbital period by a few percent
at most, the likely orbital period is 37--38 min.

\subsection{SDSS J0804}

Of all our `outbursting' AM CVn systems, SDSS J0804 shows the
  lowest amplitude variations (1.2 mag) and has the longest orbital
  period. It does not show obvious outbursts such as KL Dra or SDSS
  J0129, but shows a light curve more similar to CR Boo (which has the
  shortest orbital period of our outbursting systems). However, in
  contrast to CR Boo there are no clear low/high states, with 90
  percent of our datapoints in the range 18.2$<g<$18.8 and only 8
  percent brighter than $g$=18.2. It is therefore not clear whether
  SDSS J0804 shows genuine outbursts but rather is just more active
  than systems which do not show outbursts.

\section{The 'dip' feature}
\label{dips}

Ramsay et al. (2010) found clear evidence for a short duration decrease
in the optical flux of KL Dra around five days after the start of the
outburst. With our more extended dataset, we have identified a total
of five such dips in KL Dra suggesting they are not
transient. Moreover, similar dips are seen in our LT optical light
curves of SDSS J0926 and CP Eri (Figure \ref{mean}) and in the
recently discovered system PTF1 J0719+4858 (Levitan et al. 2011).

We show in Figure \ref{mean} the outburst light curve of KL Dra in the
optical and UV bands. Given the relatively sparse cadence of our data,
this is accurate only to within a few days. However, the dip lasts
2--3 days and has a depth of a factor of $\sim$5 compared to the
pre-dip light curve.  Although the UV data are more scattered compared
to the optical data, the UV flux tends to be weaker at the time of the
optical dip. (We have included {\sl Swift} UV data which were taken in
April 2010 -- after the paper by Ramsay et al. 2010 was accepted). Our
data do not have high enough cadence to determine whether a dip is
seen in {\it every} outburst of KL Dra.

A dip-like feature has also been reported in the AM CVn system V803
Cen by Kato et al. (2004) who noted its similarity to the 2001
super-outburst of WZ Sge. This dip has been interpreted as a
re-brightening due to a recurring thermal instability (eg Osaki et al.
2001).  However, compared to KL Dra the dip occurs much longer after
the on-set of the outburst.  We note that dips have also been seen in
some hydrogen accreting CVs, eg T Leo, (Kato 1997), which are thought
to due to a normal outburst triggering a superoutburst (van der Woerd
\& van Paradijs 1987).

\section{Outbursting systems as a function of orbital period}

One of the main motivations for our study was to test the prediction
of Smak (1983), and later elaborated by Tsugawa \& Osaki (1997), who
suggested that systems with orbital periods in the range $\sim$20--40
min would experience outbursts.  For AM CVn systems with
$P_{orb}$\gtae40 mins, the accretion disc is predicted to be cooler
than the ionisation state of helium. For binaries with
$P_{orb}$\ltae20 mins, the disc is predicted to be hotter than the
ionisation state of helium. Since these discs were expected to be
stable it was predicted that systems would not show photometric
outbursts.

Tsugawa \& Osaki (1997) indicated (their Figure 3) the predicted
location of these thermally stable and unstable regions in the
$P_{orb}$, $\dot{M}$ plane for AM CVn binaries.  Depending on whether
the donor star is fully or partially degenerate may shift a particular
source to the stable or unstable region. We have taken Figure 3 of
Tsugawa \& Osaki (1997) as a starting point, but employed the
predicted relationship between $P_{orb}$ and $\dot{M}$ from Deloye et al.
(2007) which is more physical than that of the earlier work.

\begin{figure}
\begin{center}
\setlength{\unitlength}{1cm}
\begin{picture}(8,11)
\put(-0.5,-1.5){\includegraphics{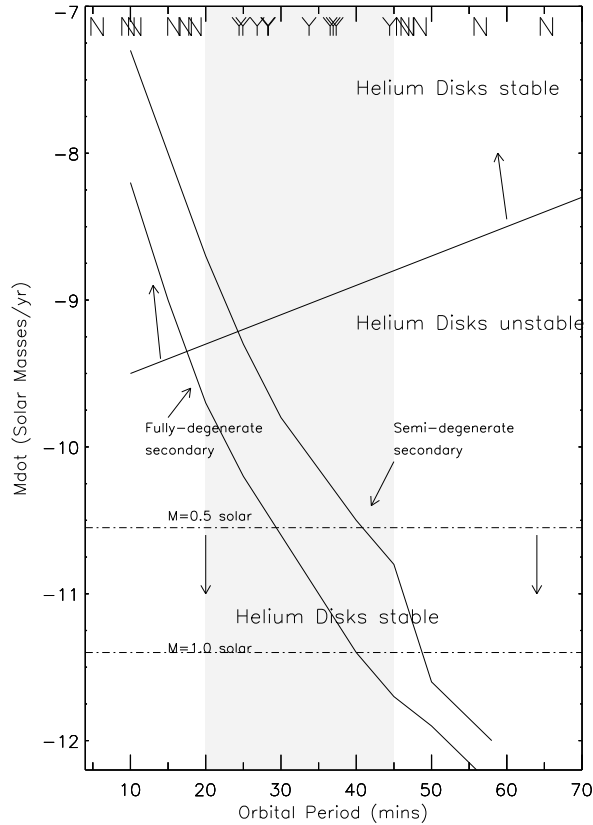}}
\end{picture}
\end{center}
\caption{We show the stable and unstable regions for AM CVn systems as
  a function of orbital period and mass transfer rate where we have
  taken the general outline from Tsugawa \& Osaki (1997) as a starting
  point but annotated and updated it. Unstable accretion disks are
  predicted to lie between the hot and the cool stable regions (top
  and bottom). The horizontal dashed-dotted lines give the location of
  the cool stable region depending on the mass of the accreting
  star. The lines travelling from the lower right to the upper left
  are the predicted mass-transfer rate as a function of orbital period
  (taken from Figure 15 of Deloye et al. 2007).  We have added a 'Y' or
  a 'N' at the appropriate orbital period depending on whether the AM
  CVn binary goes into outbursts or not.}
\label{orb-mass}
\end{figure}

In the top of Figure \ref{orb-mass} we place a `Y' or a `N' indicating
whether a source shows outbursts or not (Table 1).  We find that all
the systems with orbital periods between 24--44 mins show outbursts
and are located in the unstable region (Figure \ref{orb-mass}). 
  However, taking into account the period of the systems which do not
  show outbursts, the unstable region potentially extends between
  19--46 mins (cf Table 1). At the lower end of the unstable range,
CR Boo (which is thought to have a semi-degenerate donor star Roelofs
et al. 2007b) is just inside the unstable regime. In the case of KL
Dra, which has a virtually identical orbital period, it is unknown
whether its donor star is semi or fully degenerate. At the shortest
orbital periods, these systems would be stable whether their donor
star was fully or semi degenerate.  At longer periods ( J0804, GP Com,
J1411) these stars would be more likely to be unstable if their donor
star was semi-degenerate and/or their accreting white dwarf was
relatively low mass.  This result indicates a remarkable agreement
between the prediction of Smak (1983), Tsugawa \& Osaki (1997) and
observation.

\section{A helium-rich dwarf-nova sequence?}

Our survey to characterise the long term behaviour of AM CVn binaries
has found that 10 out of the 27 systems has shown at least one
outburst. Moreover, the outbursting systems have periods between 24
and 44 mins -- a result which is remarkably consistent with
theoretical predictions. It appears that systems with $P_{orb}$\ltae24
mins have an accretion disc which has a mean temperature always above
the ionisation state of helium, while systems with $P_{orb}$\gtae 44
mins have a disc with a mean temperature below the ionisation state of
helium.

Of the seven systems which showed at least one outburst in our LT data,
there is some evidence that the outbursts fall into three rather
distinct characteristics, the nature of which may be related (at least
partly) to their orbital period. KL Dra (25.0 min), J0926 (28.3 min)
and CP Eri (28.4 min) all show outbursts lasting a week or two with a
distinctive dip in the outburst profile. The recently discovered
system PTF1 J0719 (26.8 min, Levitan et al. 2011) also shows outbursts
lasting several weeks, but in addition exhibits bursts lasting several
days (which is very similar to what we observe in KL Dra, although
they are not well sampled).  Given our low cadence it is possible that
the 50--60 day outbursts seen in KL Dra are super-outbursts, and the
short duration bursts are normal outbursts.

CR Boo which has a period just shorter than KL Dra (24.5 min) shows a
rapid change between high and low states. In addition it has shown a
peculiar `cycling' state when it shows a prominent modulation ($\sim$1
mag) on a period of $\sim$19-22 hrs (Patterson et al. 1997). This
photometric feature has also been seen in V803 Cen (26.9 min)
(Patterson et al. 2000). At the longer period boundary, SDSS J0804
(44.5 mins) also shows enhanced levels of activity, but not regular
outbursts. This may hint that systems on the boundary between having a
stable or a non-stable accretion disc are likely to be unstable. On
the other hand, given KL Dra and CR Boo have practically identical
periods it is likely that factors other than the orbital period (for
instance, the nature and metallicity of the donor star) affects the
long term photometric behaviour.

In complete contrast, V406 Hya (33.8 min) and J0129 ($\sim$37 min)
show outbursts which are up to 5--6 mag in amplitude and which last
several months. Evidence exists that J1240 (37.4 mins) also shows high
amplitude and long duration outbursts (Shears et al. 2011).  The
similar period of these systems makes observations of J1427 (which has
a photometric period -- probably a superhump -- of 36.6 min, Shears et
al 2011) particularly interesting. Does it also show similar long
outbursts?  In the case of SDSS J2047+0008, which currently does not
have a known orbital period, the fact that it has been found to go
into outburst, strongly suggests that it has an orbital period between
24--44 mins.

\section{Summary}

Our survey of AM CVn binaries using the Liverpool Telescope coupled
with existing published data shows that $\sim$1/3 of these helium
accreting compact binaries undergo outbursts. These outbursting
systems have orbital periods within the range 24--44 mins. This result
is in excellent agreement with theoretical predictions. Moreover, we
find tentative evidence for the characteristics of these outbursts to
be dependent on orbital period. Systems at the lower end of this
period range show outbursts which last a week or two and tend to show
a characteristic dip a handful of days after the start of the
outburst. At longer periods, the outbursts have greater amplitude and
last for several months. Systems at the edge of this period range,
whether the disc is close to being in a stable hot state or a stable
cool state, show rapid changes in brightness.

This sequence can be seen in the context of the rapidly diminishing
mass transfer rate between 20--40 mins. At shorter periods the higher
mass transfer rate enables the disc to achieve a critical mass on a
shorter timescale than for longer period systems where the mass
transfer rate is significantly lower. Based on our survey we predict
that J1427 will also exhibit large amplitude long duration outbursts.
Indeed we encourage observations of J1427, V406 Hya and J0129 to
detect further outbursts so that they can be studied in much greater
detail, and at different wavelengths, and determine how similar they
are to the hydrogen accreting WZ Sge stars.

Using the models of Nelemans et al. (2001, 2004), Roelofs, Nelemans \&
Groot (2007c) predict the population surface density of AM CVn systems
as a function of Galactic latitude. For $|b|<30^{\circ}$ they predict
that around half of all AM CVn binaries are expected to have orbital
periods in the range 20--40 mins (irrespective of whether one assumes
`optimistic' or `pessimistic' assumptions for their space density),
although this fraction is lower for higher Galactic latitudes.
  Although we find the duty cycle is quite variable from source to
  source (and indeed, not especially well constrained), our results
  indicate that a duty cycle of $\sim$1/4 is not untypical, which may
  be due to the small disks in AM CVn systems.  Given this reasonably
  high duty cycle, surveys which followup blue transient systems may
  provide the best method with which to discover AM CVn systems. The
recent discovery of PTF1 J0719 using the Palomar Transient Factory
confirms such a view (Levitan et al. 2011).

\section{Acknowledgements}

The Liverpool Telescope is operated on the island of La Palma by
Liverpool John Moores University in the Spanish Observatorio del Roque
de los Muchachos of the Instituto de Astrofisica de Canarias with
financial support from the UK Science and Technology Facilities
Council. TB and DS acknowledges STFC for a PhD Studentship and an
Advanced Fellowship respectively. We thank the referee for useful
comments on the original manuscript.

\end{document}